# Diabetes prediction using Machine Learning algorithms and ontology

**Hakim El Massari, Zineb Sabouri, Sajida Mhammedi, and Noreddine Gherabi**

*National School of Applied Sciences, Sultan Moulay Slimane University, Lasti Laboratory, Khouribga, Morocco*
*{h.elmassari, n.gherabi}@usms.ma*

## Abstract.

Diabetes is one of the chronic diseases, which is increasing from year to year. The problems begin when diabetes is not detected at an early phase and diagnosed properly at the appropriate time. Different machine learning techniques, as well as ontology-based ML techniques, have recently played an important role in medical science by developing an automated system that can detect diabetes patients. This paper provides a comparative study and review of the most popular machine learning techniques and ontology-based Machine Learning classification. Various types of classification algorithms were considered namely: SVM, KNN, ANN, Naive Bayes, Logistic regression, and Decision Tree. The results are evaluated based on performance metrics like Recall, Accuracy, Precision, and F-Measure that are derived from the confusion matrix. The experimental results showed that the best accuracy goes for ontology classifiers and SVM.

## 1. INTRODUCTION

Diabetes is a group of the deadliest and metabolic diseases in which the level of blood sugar in the human body is abnormally high. It impacts the body's capacity to produce the hormone insulin. High blood sugar commonly causes many complications such as intensified thirst, increased hunger, and frequent urination if diabetes goes untreated and undiagnosed at an early stage. Therefore, early detection is the only way to avoid complications by using the trending technology for ontology-based and machine learning.

Machine learning (ML) is one of the most rapidly developing fields of computer science, with several applications. It refers to the process of extracting useful information from a large set of data. ML techniques are used in different areas such as medical diagnosis, marketing, industry, and other scientific fields. ML algorithms have been widely used in medical datasets and are best suited for medical data analysis. There are various forms of ML, including classification, regression, and clustering. Depending on the problem that we are trying to solve, each form has a distinct result and impact. In our work, we focus on classification methods, which are applied to classify a given dataset into predefined groups and to predict future activities or information to that data due to its good accuracy and performance. Classification algorithms are usually employed in the medical domain especially in diagnosing the diseases such as diabetes. Therefore, the commonly used machine learning classification [1] namely SVM, KNN, ANN, Naive Bayes, Logistic regression, and Decision Tree are applied to identify diabetes patients at an early period.

On the other side, Ontology is one of the most adopted approaches to manage, organize and extract data during the previous decades. It is a data representation method that has been successfully implemented in a variety of fields, especially the medical domain. It is important in computer science because of its capacity to represent diverse concepts and their relationships in different disciplines. In actuality, no single ontology is sufficient to follow the growing demands of today's healthcare, and the ontologies must be integrated with algorithms of machine learning to support data integration and analysis [2]. In previous work, we already created and explored an ontology-based model capable of predicting diabetes patients by using an ontology classifier based on a decision tree algorithm.

In this study, we aim to make a comparative analysis among the six popular classification techniques and ontology-based machine learning classification based on carefully chosen parameters such as Precision, Accuracy, F-Measure, and Recall, which are derived from the confusion matrix.

The organization of the remainder of the paper is as follows: Sect. 2 represents the literature review of related classification algorithms in the field of diabetes prediction. Sect. 3 we present technologies and methods used in this comparative analysis. Sect. 4 we describe the performance metrics used to evaluate the models. Sect. 5, we present results and discussion. Finally, Sect. 6 presents future work and conclusions.

## 2. RELATED WORK

Recently researchers have published a considerable amount of research to identify diabetic patients based on symptoms by applying machine-learning techniques. In [3], the authors propose a model that can predict is the patient has diabetes or not. This model is based on the prediction precision of powerful machine learning algorithms, which use certain measures such as precision, recall, and F1-measure. The authors use Pima Indian Diabetes (PIDD) dataset to predict diabetic onset based on diagnostics manner. The results obtained using Logistic Regression (LR), Naïve Bayes (NB), and K-nearest Neighbor (KNN) algorithms were 94%, 79%, and 69% respectively. In the paper [4], the authors use seven ML algorithms on the dataset to predict diabetes, they found that the model with Logistic Regression and SVM were better on diabetes prediction, they built a NN model with a different hidden layer and observed the NN with two hidden layers provided 88.6% accuracy.

The study applied in the paper [5] uses several machine learning classification algorithms (Gaussian Naive Bayes, K-Nearest Neighbors, Artificial Neural Network, Logistic

Regression, Decision Tree, Random Forest, and Support Vector Machine) on the PIID dataset. Logistic Regression got the best accuracy result.

Sarwar and al [6], discuss predictive analytics in healthcare, a number of machine learning algorithms are used in this study. For experiment purposes, a dataset of patient's medical is obtained. The performance and accuracy of the applied algorithms are discussed and compared

In the paper [7], the authors propose a diabetes prediction model for the classification of diabetes including external factors responsible for diabetes along with regular factors like Glucose, BMI, Age, Insulin, etc. Classification accuracy is improved with the novel dataset compared with existing dataset.

On a dataset of 521 instances (80 % and 20 % for training testing respectively), [8] authors applied 8 ML algorithms such as logistic regression, support vector machines-linear, and nonlinear kernel, random forest, decision tree, adaptive boosting classifier, K-nearest neighbor, and naïve bayes. According to the results, the Random Forest classifier achieved 98 % accuracy compared to the other.

In [9], the researchers used machine-learning algorithms including Logistic Regression, Gaussian Process, Adaptive Boosting (AdaBoost), Decision Tree, K-Nearest Neighbors, Multilayer Perceptron, Support Vector Machine, Bernoulli Naive Bayes, Bagging Classifier, Random Forest, and Quadratic Discriminant Analysis. The Random Forest classifier performs better and achieved a 98 % accuracy, which is higher than the other three algorithms.

To predict diabetes at an early stage, the paper [10] proposes a novel approach to diabetes prediction using significant attributes. Various tools are used to determine attribute selection for clustering and prediction. The results indicate a strong association of diabetes with body mass index (BMI) and glucose level. Several techniques for predicting diabetes are used such as artificial neural network (ANN), random forest, and K-means clustering for the prediction of diabetes, and the ANN technique provided the best accuracy

Another method is used for diabetes prediction [11]. In this method the authors propose a novel approach of machine learning algorithms applied in hadoop based clusters for diabetes prediction. This approach is applied in the Pima Indians Diabetes Database and Digestive Diseases and the results obtained show that the ML algorithms produce the best accurate diabetes predictive.

In this experimental analysis [12] four machine learning algorithms, Random Forest, K-nearest neighbor, Support Vector Machine, and Linear Discriminant Analysis are used in the predictive analysis of early-stage diabetes. High accuracy of 87.66 % goes to the Random Forest classifier.

In another way, the authors of the paper [13] have built models to predict and classify diabetes complications. In this work, several supervised classification algorithms were applied to predict and classify 8 diabetes complications. The complications include some parameters such as metabolic syndrome, dyslipidemia, nephropathy, diabetic foot, obesity, and retinopathy.

In [14], the authors present two approaches of machine learning to predict diabetes patients. Random forest algorithm for the classification approach, and XGBoost algorithm for a hybrid approach. The results show that XGBoost outperforms in terms of an accuracy rate of 74.10%.

Authors in this article [15] tested machine learning algorithms such as support vector machine, logistic regression, Decision Tree, Random Forest, gradient boost, K-nearest neighbor, Naïve Bayes algorithm. According to the results, Naïve Base and Random Forest classifiers achieved 80% accuracy compared to the other algorithms.

## 3. TECHNOLOGIES AND METHOD

The experimentation is carried out using the methods and technologies described in the next subsections. The process of developing this comparative analysis is illustrated in Figure 1.

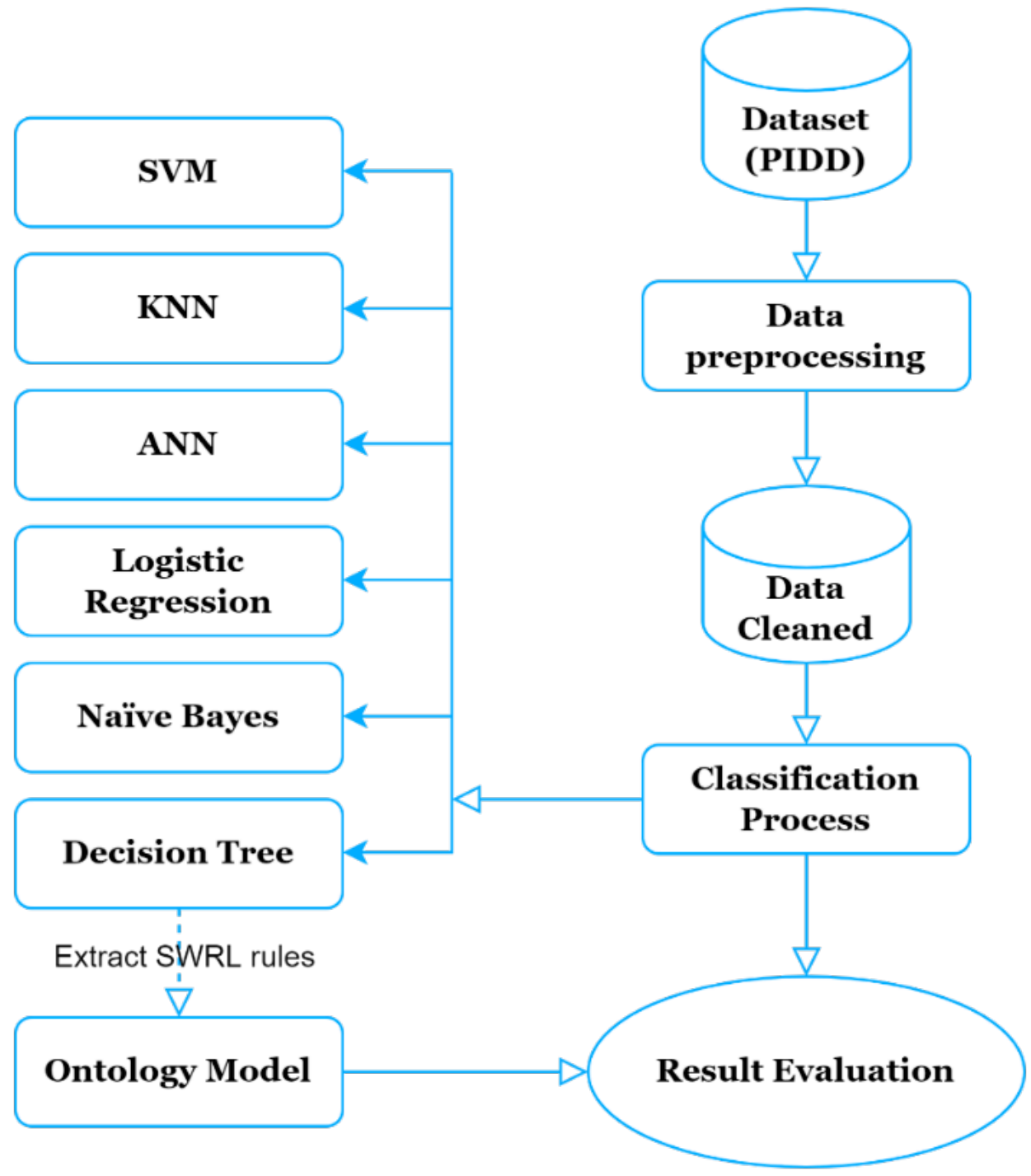

Figure 1. Experimental flowchart.

### *3.1. Dataset*

The dataset called Pima Indians Diabetes Database (PIDD) is originally from the National Institute of Diabetes and Digestive and Kidney Diseases. The purpose is to expect based on diagnostic measurements whether a patient has diabetes. It has 768 instances and 8 numerical attributes plus a class (preg, plas, pres, skin, insu, mass, pedi, age, class).

After the dataset pre-processing step using UCI Machine Learning, the output file in CSV format will be transformed into ARFF format.

### *3.2. Machine learning algorithms*

After preparing the dataset, we import it into Weka software, which contains tools for data preparation, classification [16], clustering, association rule exploration, visualization [17] and Similarity [18]. We used the six most commonly used classifiers to classify binary datasets (SVM, KNN, ANN, Logistic Regression, Naïve Bayes, Decision Tree). The results of the classifiers can be found in section 5.

### *3.3. Ontology model*

The approach used to classify the dataset using the ontology model was published and detailed in our previous work [2], we recommend reading it for more details. Here, we will give some details briefly.

The ontology was created by the open-source platform "Protégé", a free ontology editor and framework for building intelligent systems [19]. Figure 2 illustrates the graphical representation of our ontology generated by the OntoGraph plugin

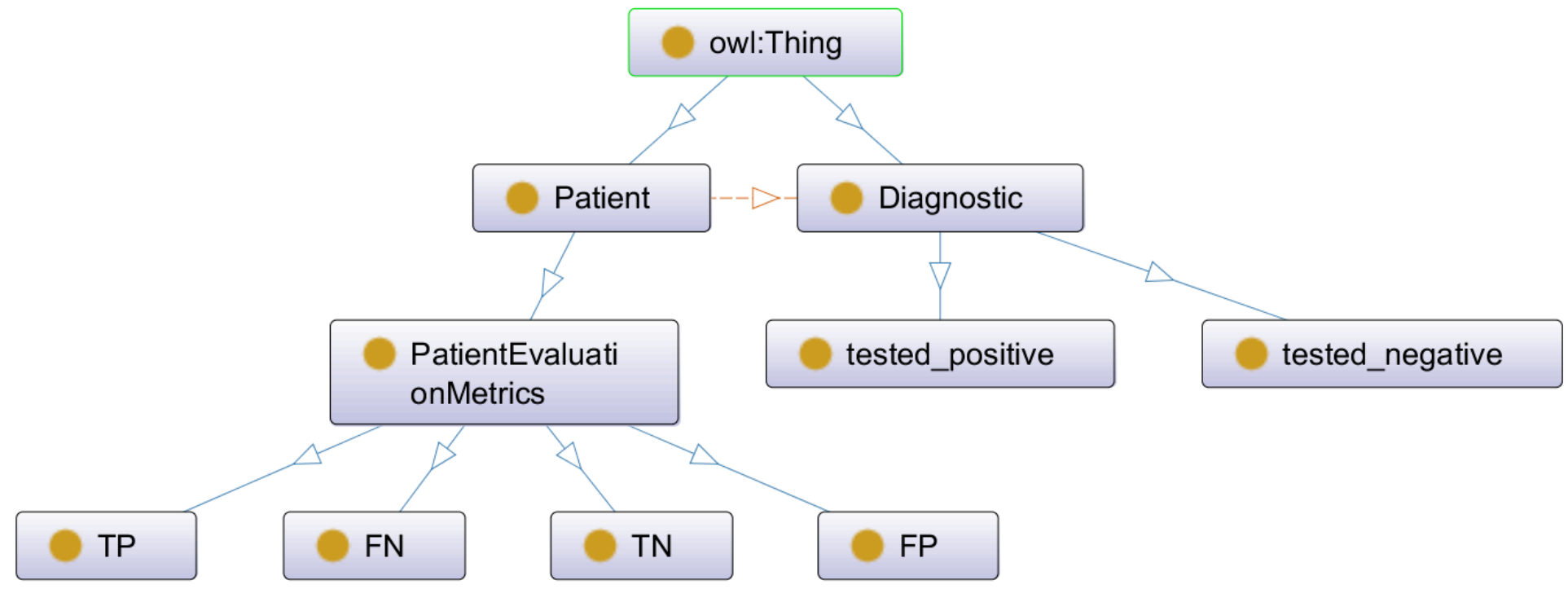

Figure 2. Graphical representation of the ontology.

The dataset is imported with the help of Cellfie, a Protégé plugin for importing spreadsheet data into OWL ontologies. Then, we extracted generated rules from the Decision Tree algorithm and import them to Protégé using the SWRLTab plugin. To execute SWRL rules and infer new ontology axioms, we used the Pellet reasoner which has a more direct functionality for working with OWL and SWRL rules. It uses the dataset and SWRL rules to induce the inference and provides the final decision where is the patient is tested negative or positive. The results of the ontology classifier are presented in Section 5.

## 4. EVALUATION

In Machine Learning, performance measurement is an essential task. It is critical to choose the right metrics to evaluate the machine learning model. Therefore, metrics are used to determine how machine learning algorithms' performance is measured and compared.

Different performance metrics are used to evaluate machine learning algorithms such as Accuracy, Precision, Recall, F-Measure, ROC Area, Kappa statistic, Root mean squared error, Root relative squared error, etc.

Almost all of the performance metrics are derived from the Confusion Matrix and the numbers inside it. The Confusion Matrix is one of the most intuitive and easiest metrics for determining the model's correctness and accuracy. It is used for classification problems with two or more types of classes as output.

The confusion matrix is a table with two dimensions ("Actual" and "Predicted"), and sets of "classes" in both dimensions. Our Actual classifications are columns and Predicted ones are Rows. For more understanding of what the confusion matrix is all about and what it represents, let's take a real example from our study where we are predicting whether a patient is having diabetes or not (1: tested positive 0: tested negative). Figure 3 illustrates the confusion Matrix details, and Table 1 describes the terms associated with the confusion matrix.

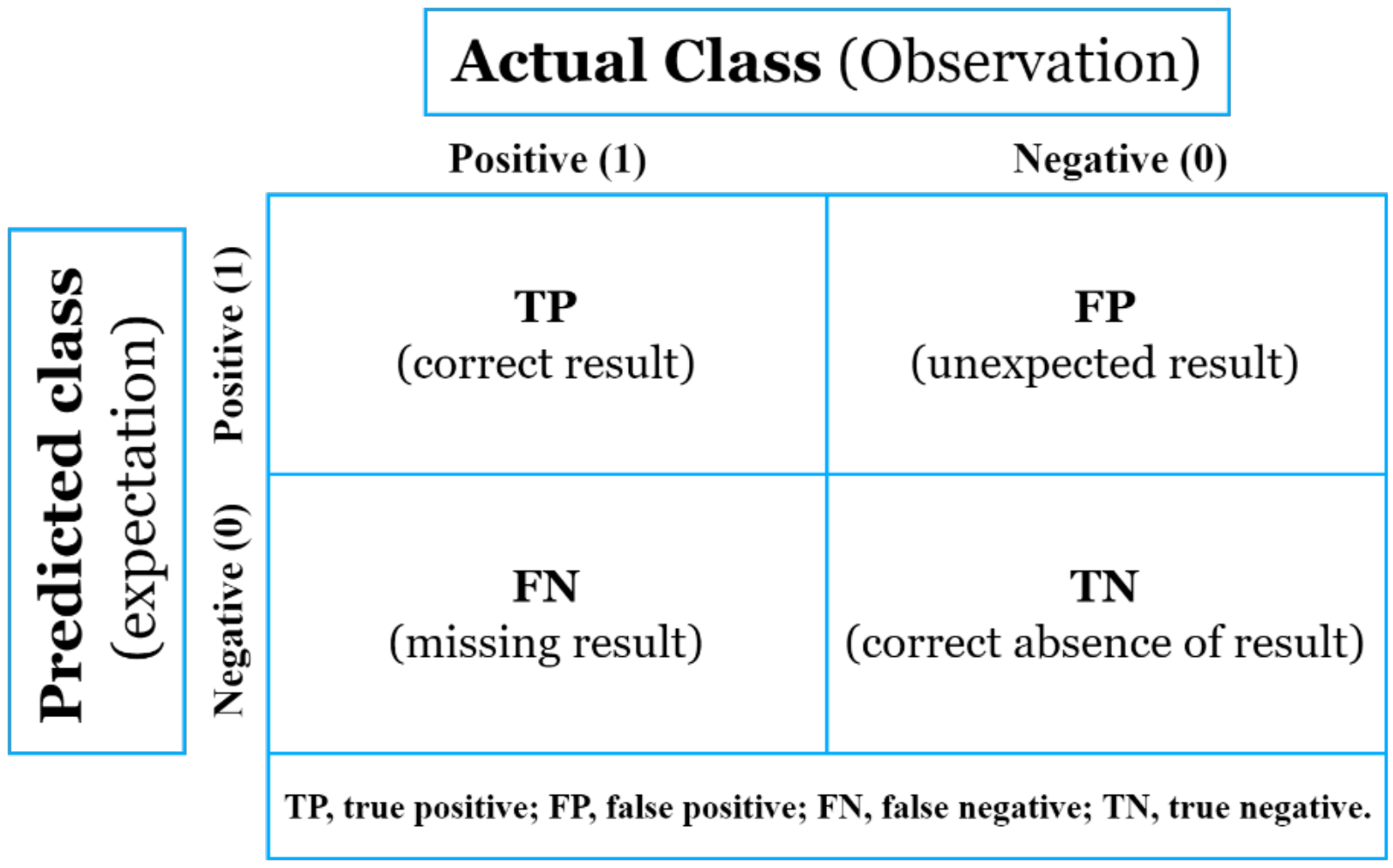

Figure 3. Confusion Matrix details.

An ideal classification performance would only have no entries for FN and FP (i.e., the number of FN equal number of FP equal zero).

**Table 1:** Terms associated with Confusion matrix

| Terms | Description |
|---|---|
| True Positives (TP) | Number of patients correctly identified as Positive |
| True Negatives (TN) | Number of patients correctly identified as Negative |
| False Positives (FP) | Number of patients incorrectly identified as Positive |

| False Negatives (FN) | Number of patients incorrectly identified as Negative |
|---|---|

Diverse measures can be derived from a confusion matrix such as Accuracy, Precision, Recall and F-Measure. The best value of accuracy, precision, and recall is 1.0, whereas the worst is 0.0. Figure 3 illustrates how to compute them from the confusion matrix.

**Accuracy (ACC):**

$$ACC = \frac{TP + TN}{TP + TN + FP + FN}$$

Accuracy is computed as the number of all correct predictions divided by the total number of the dataset, which is the number of patients that are identified correctly in total in our case.

**Precision (PREC):**

$$PREC = \frac{TP}{TP + FP}$$

PREC is computed as the number of correct positive predictions divided by the total number of positive predictions.

**Recall (REC):**

$$REC = \frac{TP}{TP + FN}$$

REC is computed as the number of correct positive predictions divided by the total number of positives. It represents the relevant patients that have been correctly detected, it is also called Sensitivity or true positive rate (TPR).

**F-Measure:**

$$\text{F-Measure} = 2 * \frac{PREC * REC}{PREC + REC}$$

F-Measure called also F-score, is a harmonic mean of precision and recall, it provides the quality of prediction.

**ROC – AUC Area:**

AUC - ROC curve is a performance measurement for the classification problems at various threshold settings. ROC is a probability curve and AUC represents the degree or measure of separability. It tells how much the model is capable of distinguishing between classes. If the value of AUC is high, the model predicts classes indicated by 0 as value 0 and classes indicated by 1 as value 1. By analogy, when the value of the AUC is high, the model is more efficient and therefore we can distinguish patients with disease and without disease.

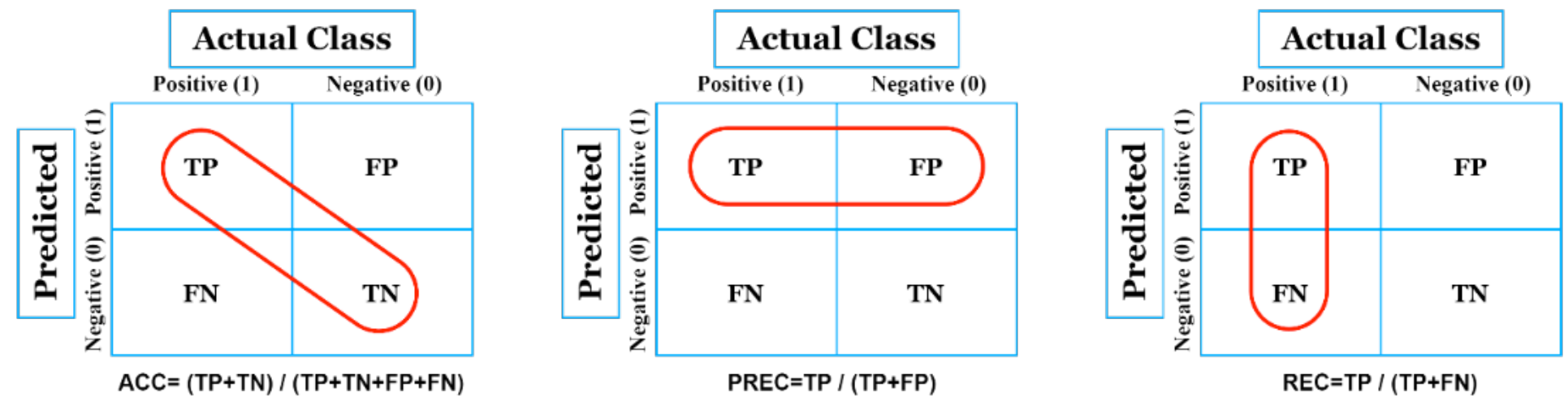

Figure 4. Performance metrics: Accuracy, Precision, Recall.

There are other metrics like Mean Squared Error (MSE), Root Mean Squared Error (RMSE), Mean Absolute Error (MAE), but generally are used in regression problems. Therefore, this comparative study will rely on the performance metrics explained above due to the dataset and algorithms used categorized in classification problems. Also, the same metrics are used to evaluate the quality of our ontology model.

In the next section, we present the result obtained from the classifiers using Weka and Protégé software.

## 5. RESULTS AND DISCUSSION

In this section, we present the result obtained from the evaluation of classifiers used in this research including the result and statistics of the ontology classifier.

This study is based on a set of criteria, on the one hand, no method applied for feature selection or performance improvement for a fair comparison of the performance of classification algorithms, on the other hand, we used two modes test: cross-validation 10 times and percentage split (split 66.0% train, remainder test) in order to enrich the study and give more visibility to these two modes.

According to the performance metrics explained in the previous section, the results of the ontology classifier are shown in Tables 2 and 3, and Figure 5. Furthermore, we present the result of Accuracy, Precision, Recall, F-Measure in Figures 6,7,8,9,10 illustrating the graphic of each metric.

Table 4 summarizes the experimental results for ML and ontology classifiers used in this study.

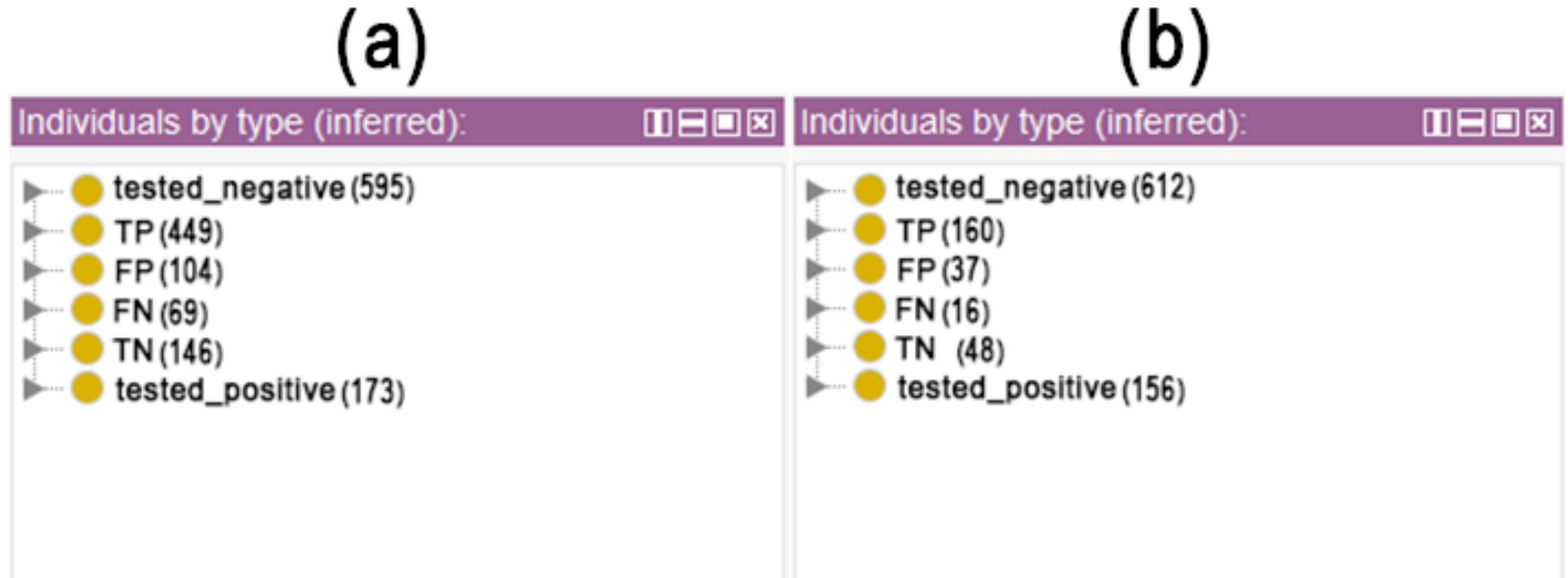

Figure 5. Statistics of inferred concepts. (a) based on 10-fold cross-validation. (b) based on 66% split mode validation.

**Table 2:** Confusion matrix of ontology classier based on 10-fold cross-validation mode.

| **Tested positive and negative classification** | | **Actual class** | |
|---|---|---|---|
| | | positive | negative |
| **Predicted class** | positive | TP : 449 | FP : 104 |
| | negative | FN : 69 | TN : 146 |

**Table 3:** Confusion matrix of ontology classier based on 66% split mode validation.

| **Tested positive and negative classification** | | **Actual class** | |
|---|---|---|---|
| | | positive | negative |
| **Predicted class** | positive | TP : 160 | FP : 37 |
| | negative | FN : 16 | TN : 48 |

- **Accuracy**

In Figure 6 and Table 4, we obtained the highest value in terms of 10-fold cross-validation mode for Ontology, SVM and Logistic Regression with 77.5%, 77.3%, 77.2% respectively. In split test mode, we obtained 80.1%, 79.7%, 79.3 for logistic regression, ontology and SVM consecutively.

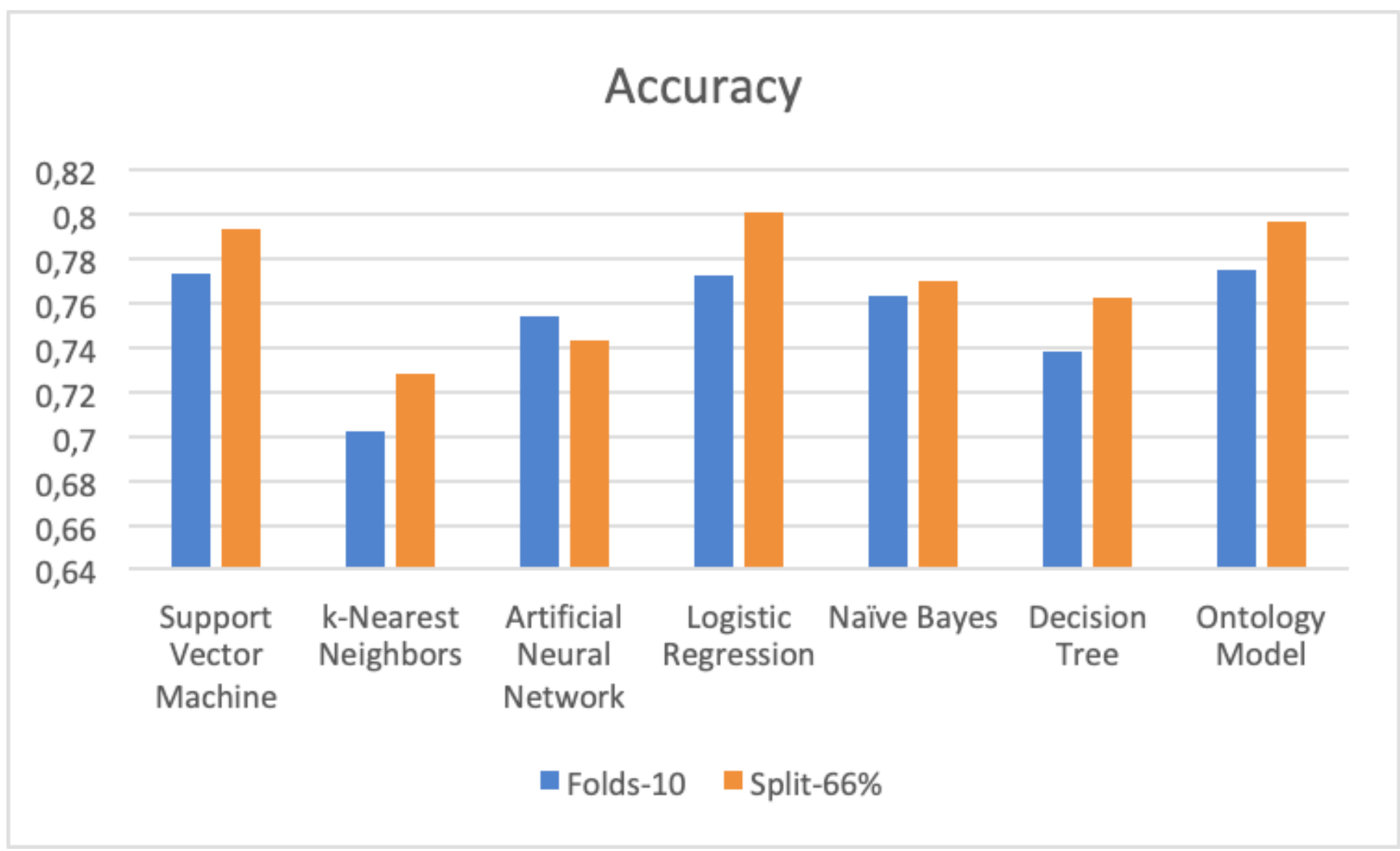

Figure 6. Comparison results of accuracy.

- **Precision**

The ontology classifier has the highest Precision of 81.2% for both test modes. Followed by Naïve Bayes and ANN. More details are shown in Table 4 and Figure 7.

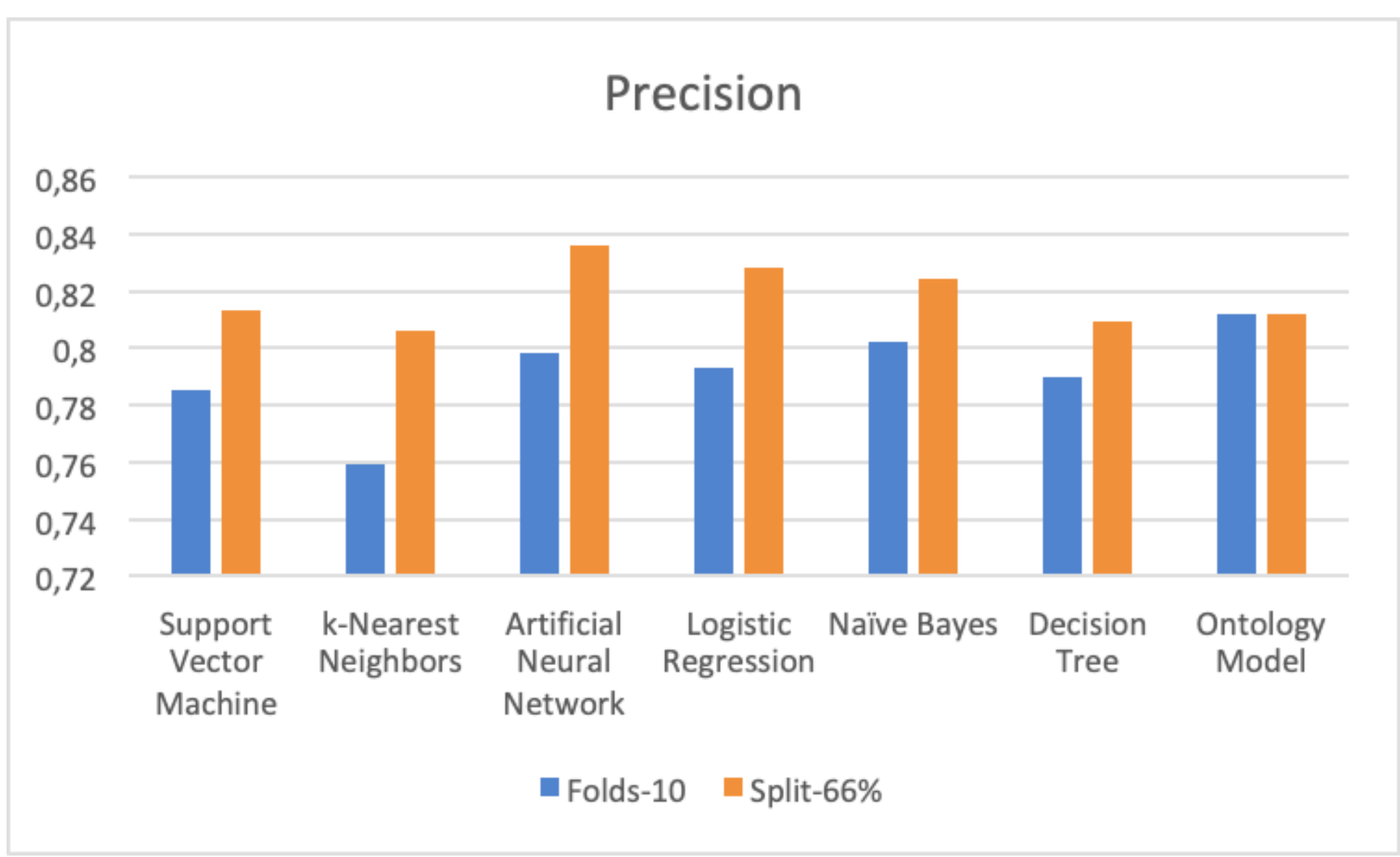

Figure 7. Comparison results of precision.

- **Recall**

From Figure 8 and Table 4, we notice that SVM had the highest value in both test modes, followed by Ontology and Logistic Regression in the last position.

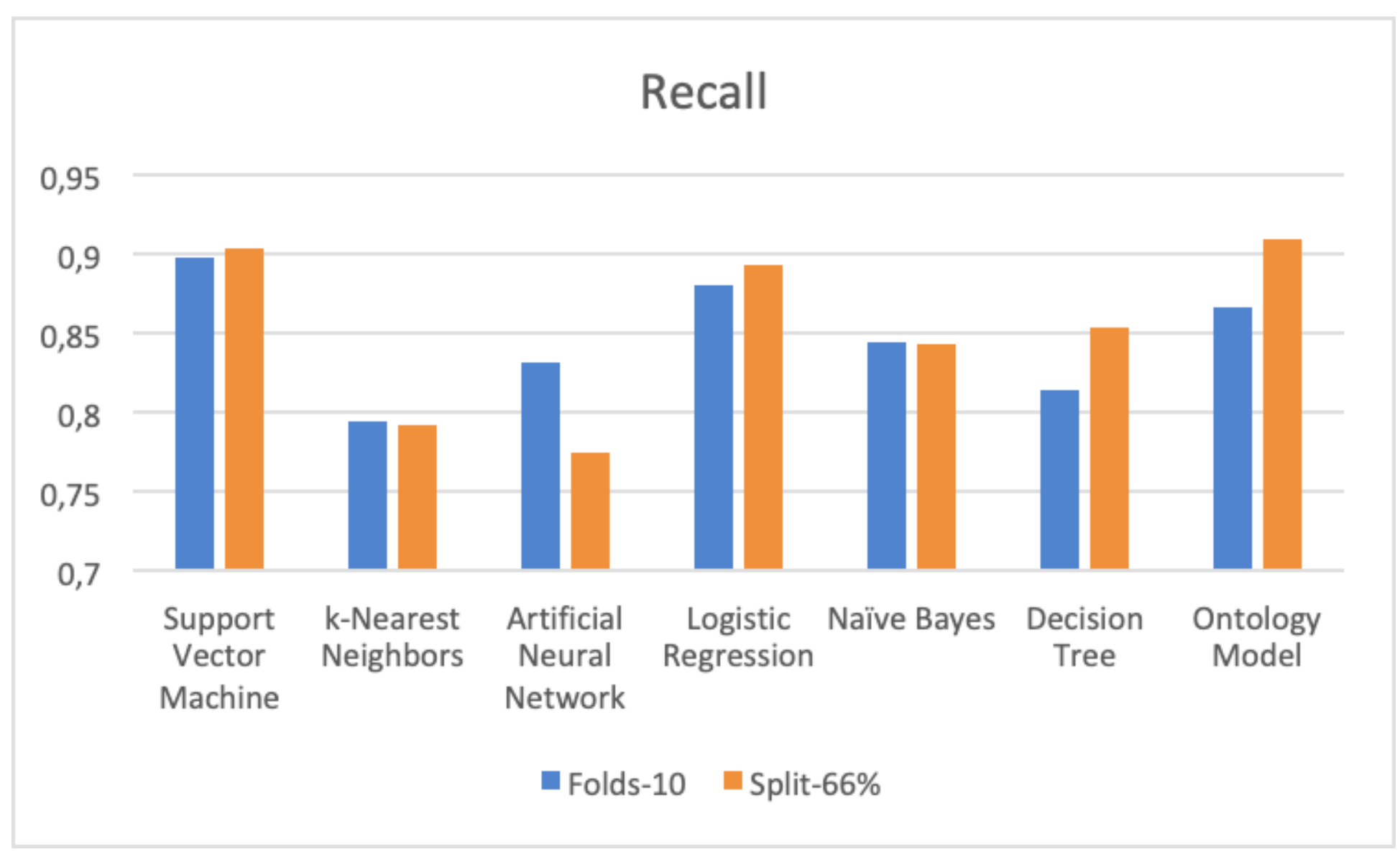

Figure 8. Comparison results of recall.

- **F-Measure**

SVM and Ontology have the same metric of F-Measure with 83.3% and ~85.8% for 10-fold cross-validation and split test mode. (See Figure 9 and Table 4)

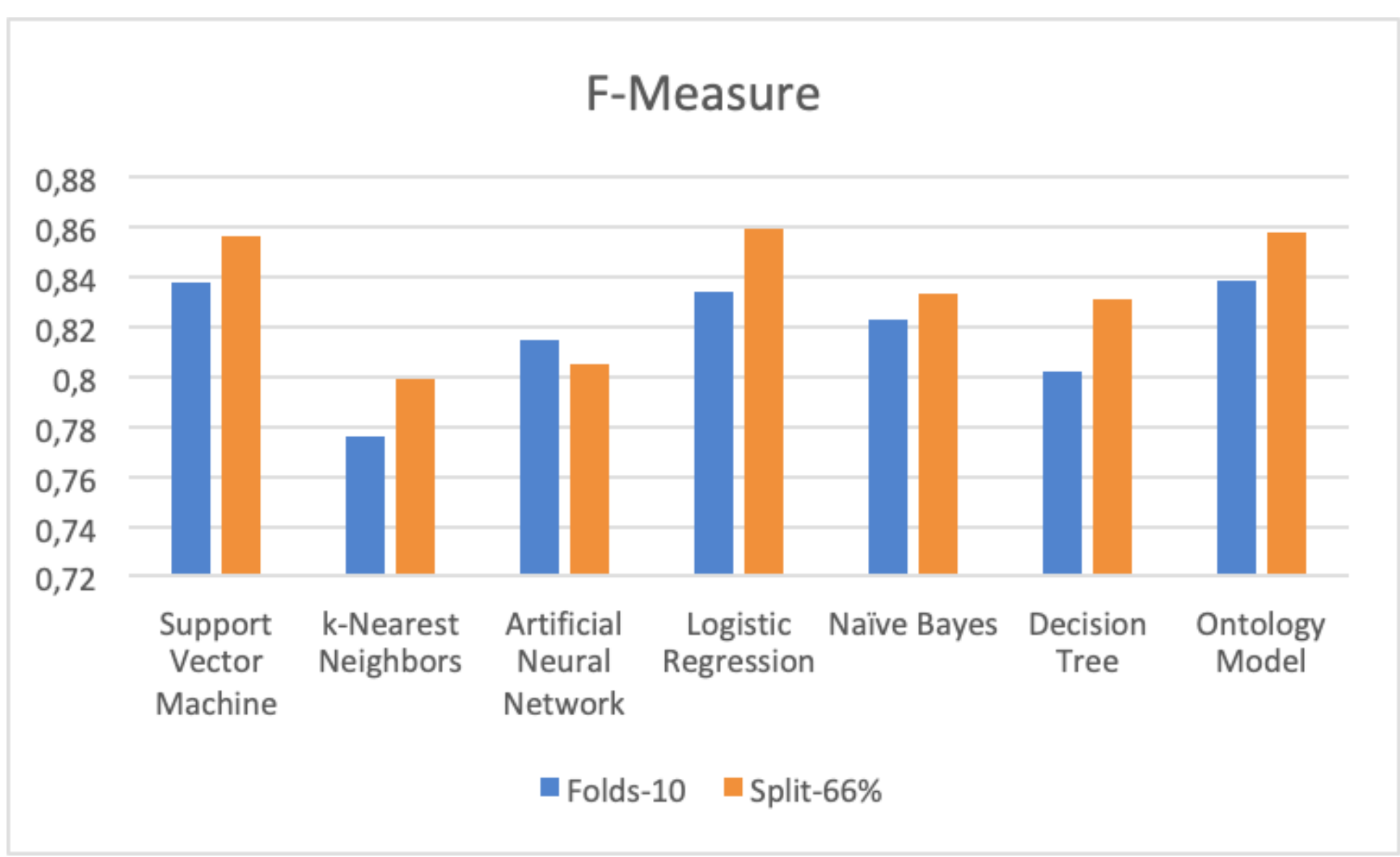

Figure 9. Comparison results of F-Measure.

- **ROC area**

Table 4 and Figure 10 show that Logistic Regression, Naïve Bayes, and Ontology have the better value of the ROC Area.

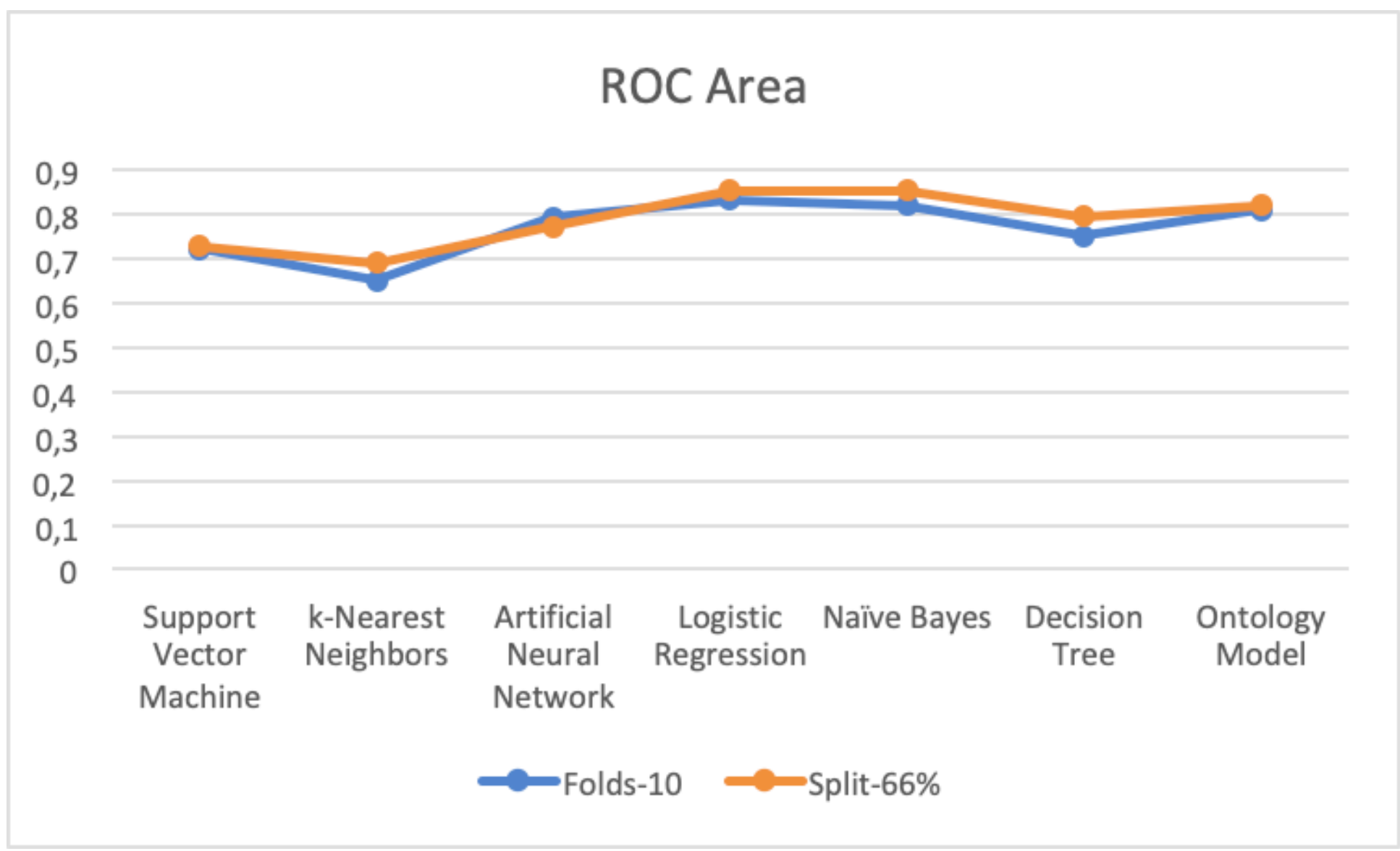

Figure 10. Comparison results of ROC area.

**Table 4:** Statistics of the experimental results for ML and ontology classifiers.

| | **Accuracy** | | **Precision** | | **Recall** | | **F-Measure** | | **ROC Area** | |
|---|---|---|---|---|---|---|---|---|---|---|
| | Folds-10 | Split-66% | Folds-10 | Split-66% | Folds-10 | Split-66% | Folds-10 | Split-66% | Folds-10 | Split-66% |
| **SVM** | 0,773 | 0,813 | 0,785 | 0,813 | 0,898 | 0,904 | 0,838 | 0,856 | 0,720 | 0,729 |
| **KNN** | 0,702 | 0,806 | 0,759 | 0,806 | 0,794 | 0,792 | 0,776 | 0,799 | 0,650 | 0,691 |
| **ANN** | 0,754 | 0,836 | 0,798 | 0,836 | 0,832 | 0,775 | 0,815 | 0,805 | 0,793 | 0,772 |
| **LR** | 0,772 | 0,828 | 0,793 | 0,828 | 0,880 | 0,893 | 0,834 | 0,859 | 0,832 | 0,855 |
| **NB** | 0,763 | 0,824 | 0,802 | 0,824 | 0,844 | 0,843 | 0,823 | 0,833 | 0,819 | 0,854 |
| **DT** | 0,738 | 0,809 | 0,790 | 0,809 | 0,814 | 0,854 | 0,802 | 0,831 | 0,751 | 0,796 |
| **Ontology** | 0,775 | 0,812 | 0,812 | 0,812 | 0,867 | 0,909 | 0,838 | 0,858 | 0,808 | 0,819 |

**Discussion:**

In our measurements, we used two test mode options, and we noticed that the percentage split was exceeded in the cross-validation test mode due to the small data mass, for this we will base by following on a cross-validation 10 times.

In this benchmarking, we used classification machine learning algorithms to retrieve the performance metrics obtained from the classifiers

We compared the ontology results to different machine learning algorithms, and the experimental results show that the ontology classifier is considered the best with a high accuracy 77.5%, followed by the SVM algorithms 77.3% and logistic regression 77.2%. We conclude that the combination of machine learning and ontological reasoning (i.e., using rules extracted from machine learning algorithms and integrating them using SWRL into the ontology) may give better results. Moreover, these comparison results confirm how the knowledge representation and reasoning capabilities of OWL ontology could provide additional benefits besides classification.

Moreover, the ontology classifier is an interpretable model, which can thus provide information on how the process makes the decision. The results of the ontology classifier are identical and comparable to those of the machine learning classifiers. The results are also human interpretable and the rules can be changed or added as needed.

Our comparative study is selective and unique in the way that we have integrated for the first-time ontology with machine learning and precisely in the field of the prediction of diabetic patients; it is therefore a first comparative analysis of ML and ontology classifiers. No meaningful comparison was made for this reason; on the other hand, researchers use different data and other methods for selection and performance improvement.

## 6. CONCLUSION AND FUTURE WORK

Machine learning techniques are widely used in all scientific fields and are responsible for revolutionizing industries across the world. The field of health has recently experienced great development in terms of the use of automatic learning mechanisms and methods. These techniques have shown effective results and could be useful in the management of chronic diseases such as diabetes.

The Semantic Web, for its part, has proven its value and strength in various fields, including the field of health, ontology as a part of the Semantic Web comes with its ability to process concepts and relationships way humans perceive interrelated concepts.

This comparative analysis summarizes the result obtained from the most common classification machine learning methods and ontology-based machine learning. The findings reveal that, even with no feature selection applied, the ontology classification method has the highest accuracy. This leads us to a new search field that we suggest and encourage researchers to contribute and create new ideas in the same context, to give more results and comparison, for the purpose of prediction, recommendation, or make a decision, etc.

From our side, we look forward to enhancing this comparative study by applying new approaches to integrate rules of machine learning with the ontology classification method, we also intend to use regression machine learning algorithms

## 7. REFERENCES

[1] Z. Sabouri, Y. Maleh, and N. Gherabi, “Benchmarking Classification Algorithms for Measuring the Performance on Maintainable Applications,” in Advances in

Information, Communication and Cybersecurity, Cham, 2022, pp. 173–179. doi: 10.1007/978-3-030-91738-8_17.

[2] H. EL Massari, S. Mhammedi, Z. Sabouri, and N. Gherabi, “Ontology-Based Machine Learning to Predict Diabetes Patients,” in Advances in Information, Communication and Cybersecurity, Cham, 2022, pp. 437–445. doi: 10.1007/978-3-030-91738-8_40.

[3] F. Alaa Khaleel and A. M. Al-Bakry, “Diagnosis of diabetes using machine learning algorithms,” Mater. Today Proc., Jul. 2021, doi: 10.1016/j.matpr.2021.07.196.

[4] J. J. Khanam and S. Y. Foo, “A comparison of machine learning algorithms for diabetes prediction,” ICT Express, vol. 7, no. 4, pp. 432–439, Dec. 2021, doi: 10.1016/j.icte.2021.02.004.

[5] P. Cıhan and H. Coşkun, “Performance Comparison of Machine Learning Models for Diabetes Prediction,” in 2021 29th Signal Processing and Communications Applications Conference (SIU), Jun. 2021, pp. 1–4. doi: 10.1109/SIU53274.2021.9477824.

[6] M. A. Sarwar, N. Kamal, W. Hamid, and M. A. Shah, “Prediction of Diabetes Using Machine Learning Algorithms in Healthcare,” in 2018 24th International Conference on Automation and Computing (ICAC), Sep. 2018, pp. 1–6. doi: 10.23919/IConAC.2018.8748992.

[7] A. Mujumdar and V. Vaidehi, “Diabetes Prediction using Machine Learning Algorithms,” Procedia Comput. Sci., vol. 165, pp. 292–299, Jan. 2019, doi: 10.1016/j.procs.2020.01.047.

[8] M. Rady, K. Moussa, M. Mostafa, A. Elbasry, Z. Ezzat, and W. Medhat, “Diabetes Prediction Using Machine Learning: A Comparative Study,” in 2021 3rd Novel Intelligent and Leading Emerging Sciences Conference (NILES), Oct. 2021, pp. 279–282. doi: 10.1109/NILES53778.2021.9600091.

[9] M. U. Emon, M. S. Keya, Md. S. Kaiser, Md. A. islam, T. Tanha, and Md. S. Zulfiker, “Primary Stage of Diabetes Prediction using Machine Learning Approaches,” in 2021 International Conference on Artificial Intelligence and Smart Systems (ICAIS), Mar. 2021, pp. 364–367. doi: 10.1109/ICAIS50930.2021.9395968.

[10] T. Mahboob Alam et al., “A model for early prediction of diabetes,” Inform. Med. Unlocked, vol. 16, p. 100204, Jan. 2019, doi: 10.1016/j.imu.2019.100204.

[11] N. Yuvaraj and K. R. SriPreethaa, “Diabetes prediction in healthcare systems using machine learning algorithms on Hadoop cluster,” Clust. Comput., vol. 22, no. 1, pp. 1–9, Jan. 2019, doi: 10.1007/s10586-017-1532-x.

[12] G. Tripathi and R. Kumar, “Early Prediction of Diabetes Mellitus Using Machine Learning,” in 2020 8th International Conference on Reliability, Infocom Technologies and Optimization (Trends and Future Directions) (ICRITO), Jun. 2020, pp. 1009–1014. doi: 10.1109/ICRITO48877.2020.9197832.

[13] Y. Jian, M. Pasquier, A. Sagahyroon, and F. Aloul, “A Machine Learning Approach to Predicting Diabetes Complications,” Healthcare, vol. 9, no. 12, Art. no. 12, Dec. 2021, doi: 10.3390/healthcare9121712.

[14] S. Barik, S. Mohanty, S. Mohanty, and D. Singh, “Analysis of Prediction Accuracy of Diabetes Using Classifier and Hybrid Machine Learning Techniques,” in Intelligent and Cloud Computing, Singapore, 2021, pp. 399–409. doi: 10.1007/978-981-15-6202-0_41.

[15] K. Pavani, P. Anjaiah, N. V. Krishna Rao, Y. Deepthi, D. Noel, and V. Lokesh, "Diabetes Prediction Using Machine Learning Techniques: A Comparative Analysis," in Energy Systems, Drives and Automations, Singapore, 2020, pp. 419–428. doi: 10.1007/978-981-15-5089-8_41.

[16] Nejjahi, R., Gherabi, N., Marzouk, A. "Towards Classification of Web Ontologies Using the Horizontal and Vertical Segmentation", Advances in Intelligent Systems and Computingthis link is disabled, 2018, 640, pp. 70–81

[17] S. Srivastava, "Weka: A Tool for Data preprocessing, Classification, Ensemble, Clustering and Association Rule Mining," Int. J. Comput. Appl., vol. 88, no. 10, pp. 26–29, Feb. 2014.

[17] Gherabi, N., Bahaj, M.: Outline matching of the 2D shapes using extracting XML data. In: Elmoataz A., et al. (eds.) ICISP 2012. LNCS 7340, pp. 502–512. Springer, Heidelberg (2012)

[18] A. Daoui, N. Gherabi, A. Marzouk: A New Approach For Measuring Semantic Similarity Of Ontology Concepts Using Dynamic Programming: Journal of Theoretical and Applied Information Technology 95 (17), 4132-4139 (2017).

[19] M. A. Musen, "The protégé project: a look back and a look forward," AI Matters, vol. 1, no. 4, pp. 4–12, Jun. 2015, doi: 10.1145/2757001.2757003.

## Biography

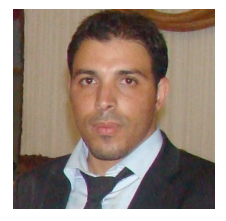

Noreddine Gherabi is a professor of computer science with industrial and academic experience. He holds a doctorate degree in computer science. In 2013, he worked as a professor of computer science at Mohamed Ben Abdellah University and since 2015 has worked as a research professor at Sultan Moulay Slimane University, Morocco. Member of the International Association of Engineers (IAENG).

Professor Gherabi having several contributions in information systems namely: big data, semantic web, pattern recognition, intelligent systems ....

He has papers (book chapters, international journals, and conferences/workshops), and edited books. He has served on executive and technical program committees and as a reviewer of numerous international conferences and journals, he convened and chaired more than 30 conferences and workshops. He is member of the editorial board of several other renowned international journals:

• Co-editor in chief (Editorial Board) in the journal "The International Journal of sports science and engineering for children" (IJSSEC).

• Associate Editor in the journal « International Journal of Engineering Research and Sports Science ».

• Reviewer in several journals / Conferences

• Excellence Award, the best innovation in science and technology 2009

His research areas include Machine Learning, Deep Learning, Big Data, Semantic Web, and Ontology

**Hakim El Massari,** Received his Ms Degree in Computer Science, from ENS Tetouan, Morocco. He worked as a visiting researcher at the Sultane Moulay Slimane University. His research areas include Machine Learning, Deep Learning, Big Data, Semantic Web, and Ontology.

**Sajida Mhammedi** Received her Ms Degree in Computer Engineering from Faculty of Science and Technologie, Beni Mellal Morocco, She worked as a visiting researcher at the Sultane Moulay Slimane University, Her research interests include Machine Learning, Semantic Web, recommendation systems, Ontology, and Big Data.

**Zineb Sabouri** Received her in Computer Engineering degree from the National School of Applied Sciences of Khouribga, Morocco. She worked as a computer engineer in a multinational. Currently, she is a Phd Student in computer science at Sultane Moulay Slimane University. Her area of interest is Machine Learning, Intelligent Systems, Deep Learning, and Big Data.